\RequirePackage{fix-cm}
\documentclass{svjour3}          
\smartqed  
\usepackage{graphicx}
\begin{document}

\title{Structural and temporal heterogeneities on networks} 

\author{Liubov Tupikina         \and
	Denis~S.~Grebenkov
}

\institute{ INSERM, Centre of Research and Interdisciplinarity, University Paris Descartes, France \at
	\email{liubov.tupikina@cri-paris.org}           
	\and
	Laboratoire de Physique de la Mati\`{e}re Condens\'{e}e (UMR 7643), \\ 
	CNRS -- Ecole Polytechnique, 91128 Palaiseau, France \at
	denis.grebenkov@polytechnique.edu
}

\date{Received: date / Accepted: date}

\maketitle

\begin{abstract}
	A heterogeneous continuous time random walk 
	is an analytical formalism for studying and modeling diffusion processes 
	in heterogeneous structures 
	on microscopic and macroscopic scales.
	In this paper we study both analytically and numerically the effects of spatio-temporal heterogeneities onto the diffusive dynamics on different types of networks. We investigate how the distribution of the first passage time is affected by the global topological network properties and heterogeneities in the distributions of the travel times. 
	In particular, we analyze transport properties of random networks and define network measures based on the first-passage characteristics. 
	The heterogeneous continuous time random walk  framework  has potential applications in biology, social and urban science, search of optimal transport properties, analysis of the effects of heterogeneities or bursts in transportation networks.
\end{abstract}

\section{Introduction}
\label{intro}

Dynamical properties of random walks on networks are related to many vital questions, such as optimality and efficiency of road systems, internet search strategy, functioning of metabolic networks \cite{montroll1965,Barthelemy,Rosenfeld}. 		
Dynamics on various types of random networks (graphs) \cite{SoodRed}  
have been extensively studied within the last decade and applied to describe epidemics and social processes \cite{BrockmanPopul}. 
Many of large networks encountered in our everyday life have the so-called small-world property
meaning that all nodes are closely connected  
\cite{Agliari2009,Zhang,Albert2001}.
Moreover, it has been  found that topological, geometric, and hydrological characteristics of a  network system 	are directly linked to the characteristics of random walks on this network \cite{Redner2002,Havlin}, such as the \emph{first passage time (FPT)}, i.e. the time of the first arrival of the random walk to a target node \cite{Bollt}. 
The first passage time characteristics are related to trapping problems, which play an important role in the control theory. Numerous problems on dynamical processes on networks \cite{Hwang2012} include finding an appropriate placing of the trapping site in a network in order to obtain the needed trapping efficiency,  search strategies etc. \cite{Lambiotte2015}. 
For homogeneous networks these problems have been studied extensively \cite{Brockman,Hudghes}, for example, using the underlying backward equation approach \cite{Bollt}.
However, the
analytical approaches, describing random walks on feature-rich (when nodes or links are endowed with some attributes \cite{Boldi}) heterogeneous networks, are lacking \cite{Bollt}.
It has been demonstrated that in continuous domains the
mean FPT can differ from the most probable FPT 
\cite{Godec,Greb2018}, in particular, moments of FPT do not fully reflect the transport efficiency. 
Moreover, 
since real-world networks are highly heterogeneous, reliable methods for analyzing transport properties on them are required \cite{CollizaTempo}. 
To our knowledge, the complete description of the first passage quantities on various networks with structural or temporal heterogeneities has not been achieved.
There is a particular reason for studying FPT properties on random networks.
Some random networks  have shown clear advantages  over other network topologies (such as complete graphs) in terms of transport optimality 
and  scaling efficiency, quantities which characterize the growth of the FPT density with the increase of network size and network diameter  \cite{Bollt}.
Recently dynamics on random networks has been studied  
using the spectral and mean-field theories \cite{Timme2012,Mighem}, which are not always applicable to heterogeneous networks.

In this article we consider  Heterogeneous Continuous Time Random Walk (HCTRW) model, in which  heterogeneity is understood from both structural and temporal perspectives  \cite{GrebTupikina}. We investigate HCTRW on various graphs  to see the interplay between the dynamical random walk characteristics and structural properties of regular and random networks.  
The structure of the paper is the following.
After presenting the model in subsection \ref{sec_mod}, we investigate in  Section \ref{sec_num_res} the first passage time of HCTRW on regular and irregular  networks with structural and temporal heterogeneities (or perturbations). 
In Sections \ref{sec_disc} and \ref{sec_conc} we discuss  the results and conclude with the main findings.  

\begin{figure}
	\includegraphics[width=3.5in]{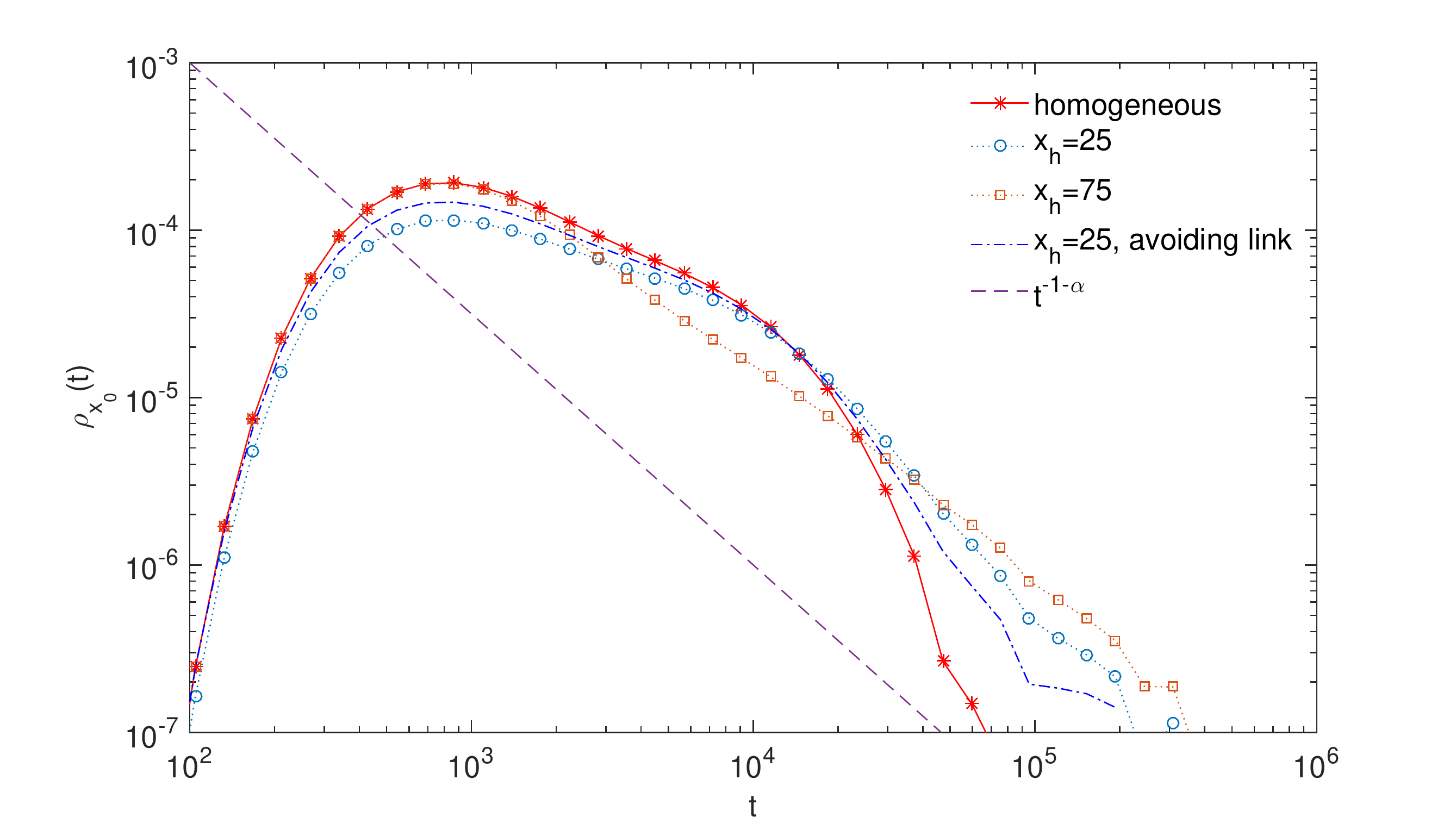}
	\caption{FPT density on a chain graph with $N=100$ nodes,   $\tilde{\psi}(s) = 1/(s\tau +1), \tau=1$ in all nodes except the trap nodes either at $x_h = 25$, or at $x_h = 75$, at which $\tilde{\psi}_{x_h}(s) = 1/(s^\alpha\tau^\alpha +1), \alpha=0.5$,  $x_0=50$ and $x_{a}=1$. The additional link  is placed around the trap $x_h=25$.  } 
	\label{fig_heter_avoid}
\end{figure}

\subsection{Model}
\label{sec_mod}
In HCTRW model \cite{GrebTupikina} a random walk moves on a graph $G$ in continuous time, jumping from one node
to another with a transition (stochastic) matrix $Q$ whose element
$Q_{xx'}$ is the probability of jumping from the node $x$ to $x'$ via link $e_{xx'}$ and the travel time needed to move from $x$ to $x'$ is a random variable drawn from the probability density $\psi_{xx'}(t)$.
The propagator of HCTRW in the Laplace domain was obtained as
\begin{equation}
\tilde{P}_{x_0x}(s) =
\frac{1-\sum_{x'}\tilde{Q}_{xx'}(s)}{s} [(I-\tilde{Q}(s))^{-1}]_{x_0x},
\end{equation}
where $\tilde{Q}(s)$ is the Laplace transform of $Q(t)$, the generalized transition matrix with elements $Q_{xx'}(t)$, which are  equal $Q_{xx'}\psi_{xx'}(t)$  \cite{GrebTupikina}.
From this formula other important quantities of the process can be derived. In particular, the probability density $\rho_{x_0}(t)$ of the first passage time to a single absorbing node $x_a$ 
can be obtained using the renewal approach
\begin{equation}
\tilde{\rho}_{x_0}(s) = \frac{\tilde{P}_{x_0x_{a}}(s)}{\tilde{P}_{x_{a}x_{a}}(s)}.
\label{eq_alt}
\end{equation} 
When there are many absorbing nodes $x_{a}$, Eq.~(\ref{eq_alt}) is not applicable but there 
are other ways to compute FPT density.

For each considered network (see below) we construct the transition matrix $Q$ and the generalized matrix $Q(t)$, given the set of the travel time distributions $\psi_{xx'}(t)$ for each link $e_{xx'}$.
Then we perform the numerical computation of the inverse Laplace transform of $\tilde{\rho}_{x_0}(s)$ by Talbot algorithm to get the FPT density $\rho_{x_0}(t)$ in time domain \cite{Talbot}.

\section{Results }
\label{sec_num_res}
We investigate the behavior of the FPT probability density on several networks. 
In each of these networks, we will select an absorbing node $x_a$ and calculate the probability density $\rho_{x_0}(t)$ of the FPT to this node from a prescribed starting point $x_0$. We will focus on the effects of heterogeneity by introducing a trap node $x_h$, at which the travel time distribution is different from the remaining nodes. Although the HCTRW model allows for arbitrary travel time distributions, we restrict our analysis to the most common exponential distribution, for which $\tilde{\psi}_{xx'}(s) = 1/(1 + s\tau)$, where $\tau$ is the mean travel time, $\tilde{\psi}_{xx'}(s)$ Laplace transform of $\psi_{xx'}(t)$. When a random walk arrives at the trap node, it is kept there for much longer times than $\tau$. This mechanism can be realized by using the exponential distribution with a much larger mean travel time. However, to highlight to effect of the trap node (heterogeneity) $x_h$, we will consider the distribution with infinite mean travel time, typically of the form $\tilde{\psi}_{x_hx'}(s) = \tilde{\psi}_{x_h}(s) =1/(1 + s^\alpha \tau^\alpha)$, with an exponent $\alpha$ between $0$ and $1$. We will investigate how the presence of such a trap node and of eventual links avoiding this trap affect the FPT density.
We start with the HCTRW on regular structures, such as one-dimensional lattice chain and finite fractals, then we study FPT properties on  random networks. 

\subsection{HCTRW on regular graphs}
\label{sec_trees}

We use the following setup of the HCTRW model on regular loopless graphs.
In a given graph we also induce an avoiding link $e_{x_h+1 x_h-1}$, which connects the nodes $x_h+ 1$ and $x_h- 1$ around the trap node $x_h$.   
Then we compute FPT densities for the cases with and without the avoiding link. 

\begin{figure}
	\includegraphics[width=2.5in]{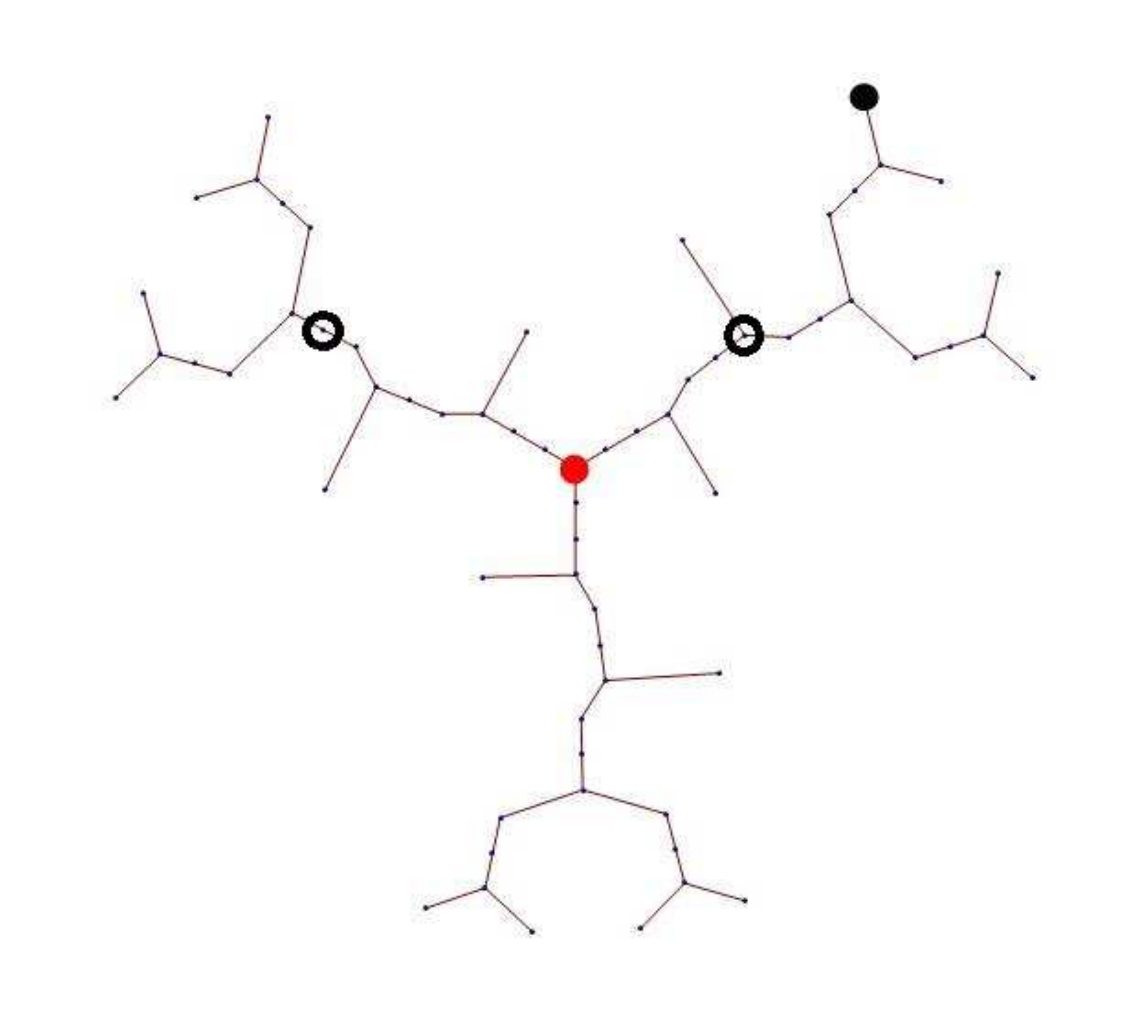}
	\\
	\includegraphics[width=3.5in]{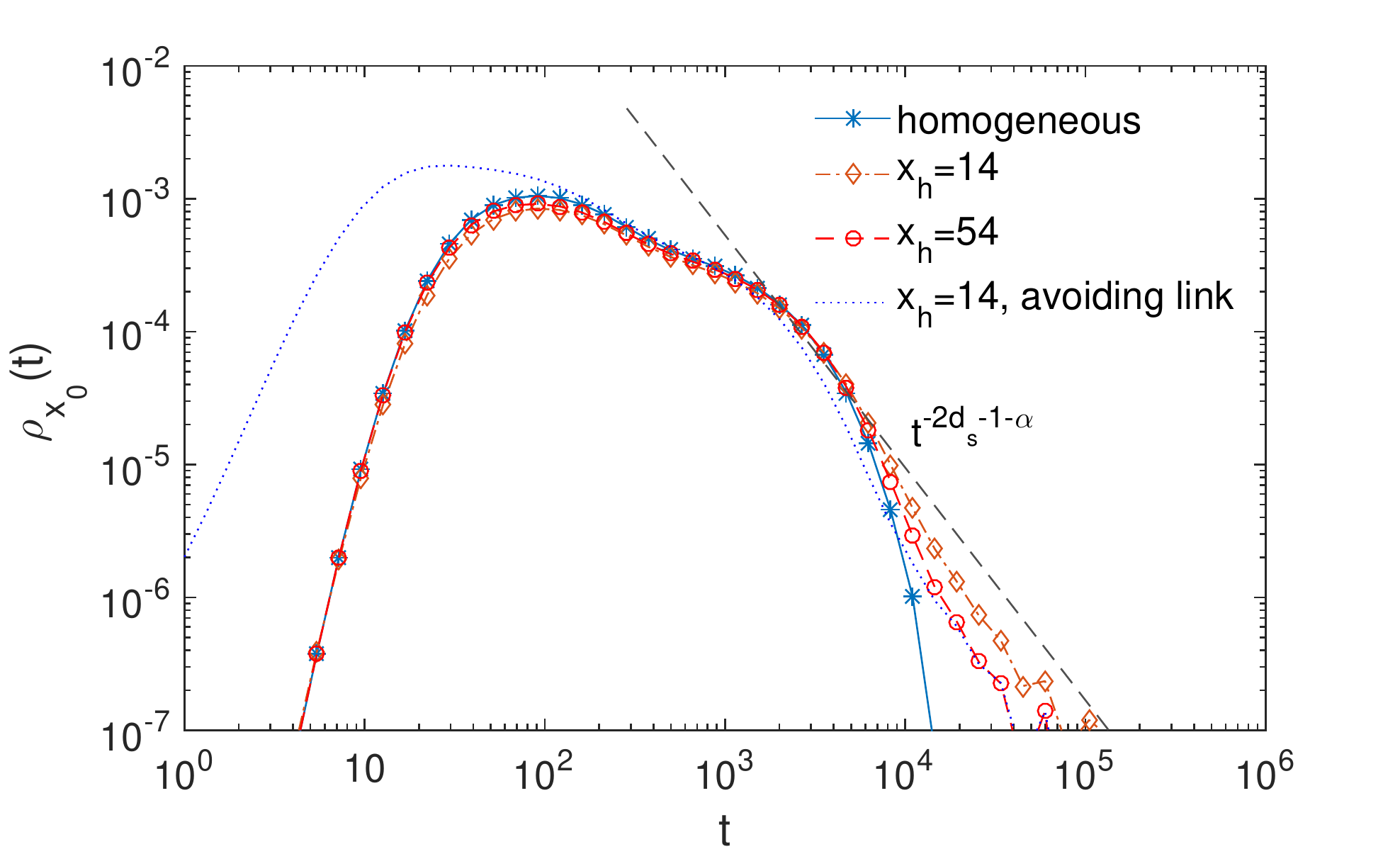}
\caption{ (Top)
		Vicsek fractal $G_{V33}$ with $N=64$ nodes.  The starting node $x_0$ is in the center (red circle), the absorbing node $x_{a}$ is one of the dead-ends (black circle), traps $x_h$ (empty circles). (Bottom)
		FPT densities on this Vicsek fractal with   $\tilde{\psi}(s) = 1/(s\tau +1), \tau=1$ in all nodes except $x_h$, at which  $\tilde{\psi}(s) = 1/(s^\alpha\tau^\alpha +1), \alpha=0.5$. 
		The trap $x_h$ is placed either on the shortest path between $x_0$ and $x_{a}$ ($x_h=14$), or outside ($x_h=54$). We also consider the case with an
		 additional link avoiding the trap $x_h=14$, which corresponds to the transition matrix with a local perturbation.}
	\label{fig_heter_avoid_fract}
\end{figure}

First, 
we consider the HCTRW on a one-dimensional chain with one absorbing node, one reflecting node and two trap nodes either to the left or to the right from $x_0$. The corresponding
FPT densities  (see Fig.~\ref{fig_heter_avoid}) 
show the same  long-time behavior $t^{-1-\alpha}$, discussed in \cite{GrebTupikina}.  
This result can be understood from the fact that a random walk on a one-dimensional chain with or without avoiding link will eventually get into a trap node even though an avoiding link potentially prevents a random walk from getting into a trap. 
As an example of regular loopless structures, we consider a generalized Vicsek fractal, $G_{Vgf}$, which resembles the dendrimers construction \cite{Blumen2004,Dolgushev}. The fractal $G_{Vgf}$ is constructed 
iteratively  in a deterministic way  by going from generation $g$ to generation $g + 1$ with coordination number $f$. 
A Vicsek fractal graph for $f=3$ and $ g=3$ is shown in  Fig.~\ref{fig_heter_avoid_fract} (top). 

We use the HCTRW setup, where all travel time distributions are   $\tilde{\psi}(s) = 1/(1+s\tau)$, except $x_h$, at which $\tilde{\psi}_{x_h}(s) = 1/(1+s^{\alpha}\tau^{\alpha})$ with $\alpha=0.5$.  
We compute the FPT density on a Vicsek fractal for two cases: when the trap is placed on the shortest path between $x_0$ and $x_{a}$ and outside this path. 
As we observe from 
Fig.~\ref{fig_heter_avoid_fract} (bottom) 
the short-time regime of the FPT density with avoiding link differs from that without avoiding link, whereas these cases give the same power-law scaling $t^{-2d_s-1-\alpha}$ in the long-time regime, where $d_s$ is the spectral dimension of the Viscek fractal.
At the same time, 
it is known that the spectral properties of graph Laplacians are reflected in dynamical properties of random walks on them. 
In the case of the Vicsek fractal, the graph Laplacian obeys the simple scaling, determined by the spectral dimension of the Vicsek fractal:   
$d_s =\frac{\ln(f+1)}{\ln(3f+3)} \approx0.557$.

\subsection{HCTRW on random networks}
\label{sec_scalfr}

Here we consider two classes of random networks: Scale-Free (SF) and Watts-Strogatz (WS) model. We construct SF networks using the preferential attachment Barabasi-Albert model  $G(N,m,m_0)$, where $N$ is the number of nodes, $m$ is the number of initially placed nodes in a network and $m_0$ is the number of nodes, a newly added node is connected to \cite{Albert2001}. The preferential attachment mechanism drives the network degree distribution to obey a power law decay with the exponent $\gamma \approx 3$ \cite{Hofstadt}. 
The small-world  Watts-Strogatz network model is denoted by $G(k,\beta)$, where $k$ is the average network degree and $\beta$ is the rewiring parameter \cite{Watts1998}. WS model is constructed from 
a regular ring lattice, a graph where each nodes is connected to $k$ neighbors, $k/2$ on each side, where each link is then rewired with the probability $\beta$.

We numerically compute the FPT density on SF networks. 
In this section we demonstrate the FPT densities on one random realization of SF network, while 
we also checked that results are qualitatively the same when considering an ensemble of random networks.

First, we investigate the influence of the individually placed traps. 
For this we randomly fix $x_0$, $x_{a}$ and place $x_h$ into two different communities of the network, i.e. groups of nodes, which are interlinked with each other more than with other nodes \cite{Barthelemy}:
(a) a trap $x_{h}$ is placed on a path (maybe not the shortest) between $x_0$ and $x_{a}$, denoted as $x_0<x_{h}<x_{a}$, hence a random walk cannot avoid getting into the trap $x_h$; 
(b) $x_{h}$ is placed in another community in respect to $x_0$ node, such that a random walk is able to reach $x_{a}$ without passing through $x_{h}$. 
Figure \ref{fig_heter0} illustrates these cases. 
The FPT densities for these two cases are compared with the FPT density for the homogeneous HCTRW, when all travel time distributions are exponential with the same parameter $\tau$: $\tilde{\psi}(s) = 1/(1+s\tau)$.  
Note that the case (b) is the homogeneous case in respect to a node $x_a$, since the trap node $x_h$ is not reached by a random walk. At the same time, 
placing a trap node in the same community with $x_0$ and $x_{a}$ changes the long-time behavior of the FPT density in comparison to the homogeneous case (Fig.~\ref{fig_heter0} (bottom)). 
In contrast, for  the case (b) the long-term behavior is the same as  for the homogeneous case.   
We note here that, communities structure  in larger networks can be less obvious than in the scheme in Fig.~\ref{fig_heter0}, and the HCTRW  framework with 
placing trap nodes in different parts of a network
can be used for analysis of communities and trapping efficiency. 

\subsubsection{Short-time regimes of FPT density on scale-free networks}
\label{subsec_sf}

From our numerical results for loopless network (subsection \ref{sec_trees})  we observe that
the FPT densities can differ from each other in the short-time regime. 
We numerically calculate the FPT density for the 
HCTRW separately in two cases of SF networks: for SF with locally tree-like structure ($m=1$ parameter of SF model) and with loops ($m>1$). 
First we test the influence of the global topological properties on the FPT density and plot it for SF networks with fixed $m_0$ and various $m$ parameters and homogeneous exponential travel times. Figure \ref{fig_SFnets_diff_m} shows that the case for $m=1$ affects mostly the short-time and intermediate-time regimes, while cases with $m>1$ slightly differ in the short-time regimes. 

Then we fix all travel times to be independent identical exponentially distributed $\tilde{\psi}(s)=1/(s\tau +1),\tau=1$, except in $x_h$, at which $\tilde{\psi}_{x_h}(s)=1/(s^\alpha\tau^\alpha +1)^\nu$, $\alpha\in(0,1)$, $\nu=2$, where an auxiliary
 parameter $\nu$ allows one to scatter more  $\rho_{x_0}(t)$ for different $x_0$. 
The FPT densities are computed for different starting points $x_0$ and fixed $x_{a}$ on a SF network (Figs.~\ref{fig_hetermany}, \ref{fig_scalefreenontree}).  
Knowing that there are different shortest path distances between $x_0$ and $x_a$, we estimate numerically the lengths of all possible shortest lengths  $|x_0,x_{a}|$ for fixed $x_{a}$ using  using Dijkstra's algorithm.  
We further explore the relation between the short-time regime of the FPT density and the intrinsic network metrics, defined by the shortest paths between nodes. For SF network with loops  there exist several shortest paths between randomly taken nodes. For example, in a SF network with $m=5, m_0=6, N=100$ three different lengths of the shortest path  exist.
Correspondingly, we clearly observe three different 
early time behaviors for the FPT density, while for large time the FPT densities behave similarly for different $x_0$. This illustrates the fact that the long-time behavior of the FPT density does not capture some important information about the system. 

Next we compare the FPT densities of SF tree and non-tree networks  for different starting positions $x_0$ (Fig.~\ref{fig_hetermany} and Fig.~\ref{fig_scalefreenontree}). 
We observe that
changing $x_0$ in a scale-free tree (Fig.~\ref{fig_hetermany})
generally affects the whole FPT density, while in a scale-free network with $m=5$ only the short-time regime is affected. This illustrates the fragility of the transport properties on tree-like structures. One of simple explanations  is that 
for the same number of nodes $N$ a 
tree network has larger variety of distances between nodes than in the case of a network with loops. 
Both, scale-free networks with and without loops obey the small-world effect. Further we study the effects of the small-world property on the FPT density.

\subsubsection{FPT density on Watts-Strogatz network}

Another example of a network with small-world property is the WS network model \cite{Watts1998}. 
The average shortest path length in the WS model gives an estimate of the small-world property. 
For $\beta=0$ ($k$-circular graph) the average shortest path length is $ N/4k$  \cite{Barrat,NewmanMoore}, while for  $\beta\neq 0$ the average shortest path length is smaller than $N/4k$ and for $b=0.2, k=8$ gives $M\approx 4$,   Fig.~\ref{fig_sw} (top). 
The FPT density for a WS network with various initial positions $x_0$ and fixed $x_{a}$  are shown in Fig.~\ref{fig_sw} (bottom).
Here we consider the WS network with the rewiring probability $\beta =0.2$, the average degree $k = 8$ and $ N=100$ nodes.  
The first observation about the short-time regime is that we clearly see $m$ distinct groups, where $m$ is the number of different shortest paths lengths in a network. 
Another observation is absence of a plateau in the intermediate regime, Fig.~\ref{fig_sw}. For SF networks with $m=5$ the intermediate regime has a characteristic plateau for $t\in[10^1,10^3]$, Fig.~\ref{fig_scalefreenontree}, while for WS networks the plateau is not found. 
In section \ref{sec_disc} we discuss in more details how one can compare FPT densities for different networks.

\begin{figure}
	\includegraphics[width=2.75in]{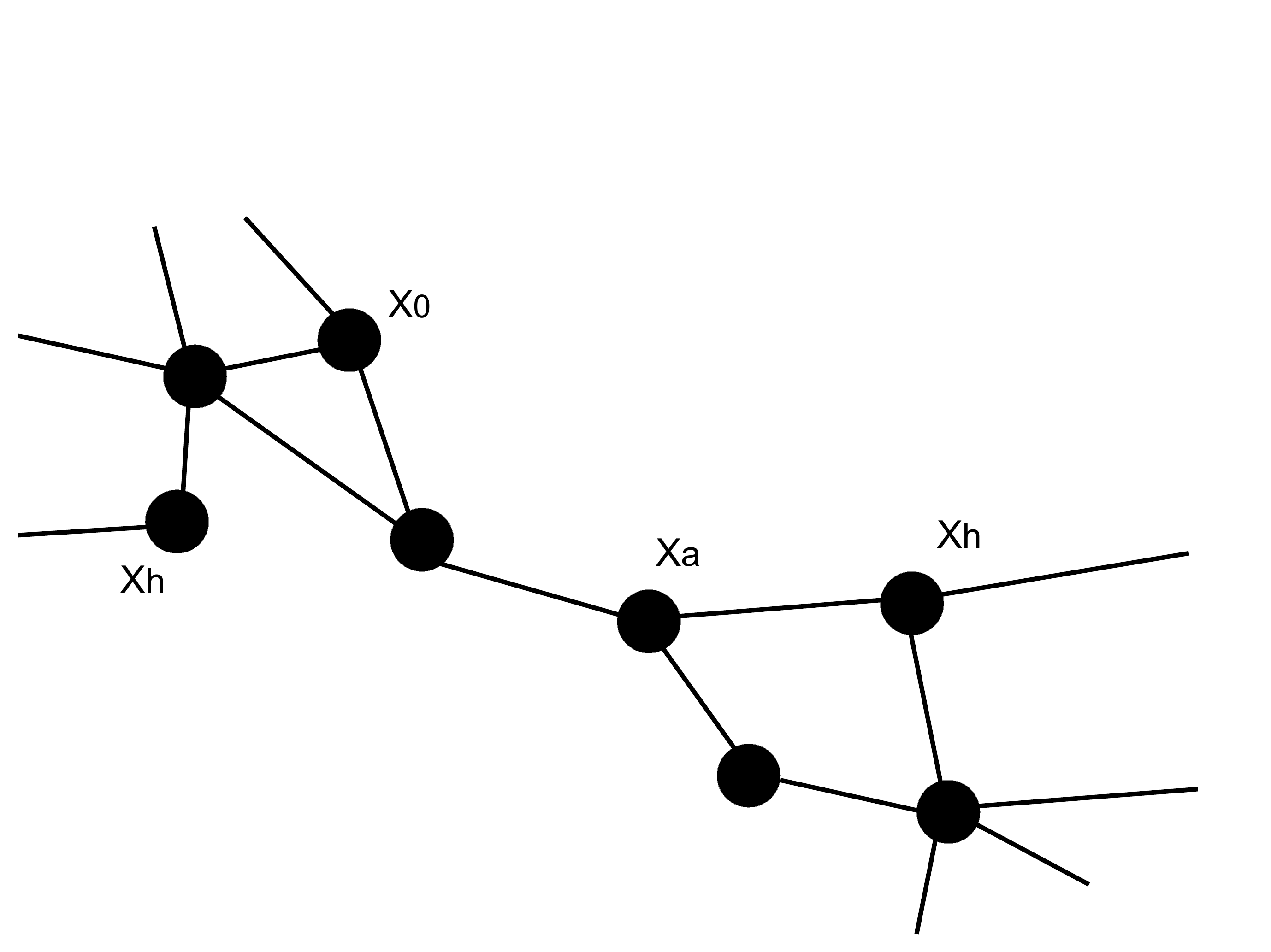}
	\\
	\includegraphics[width=3.5in]{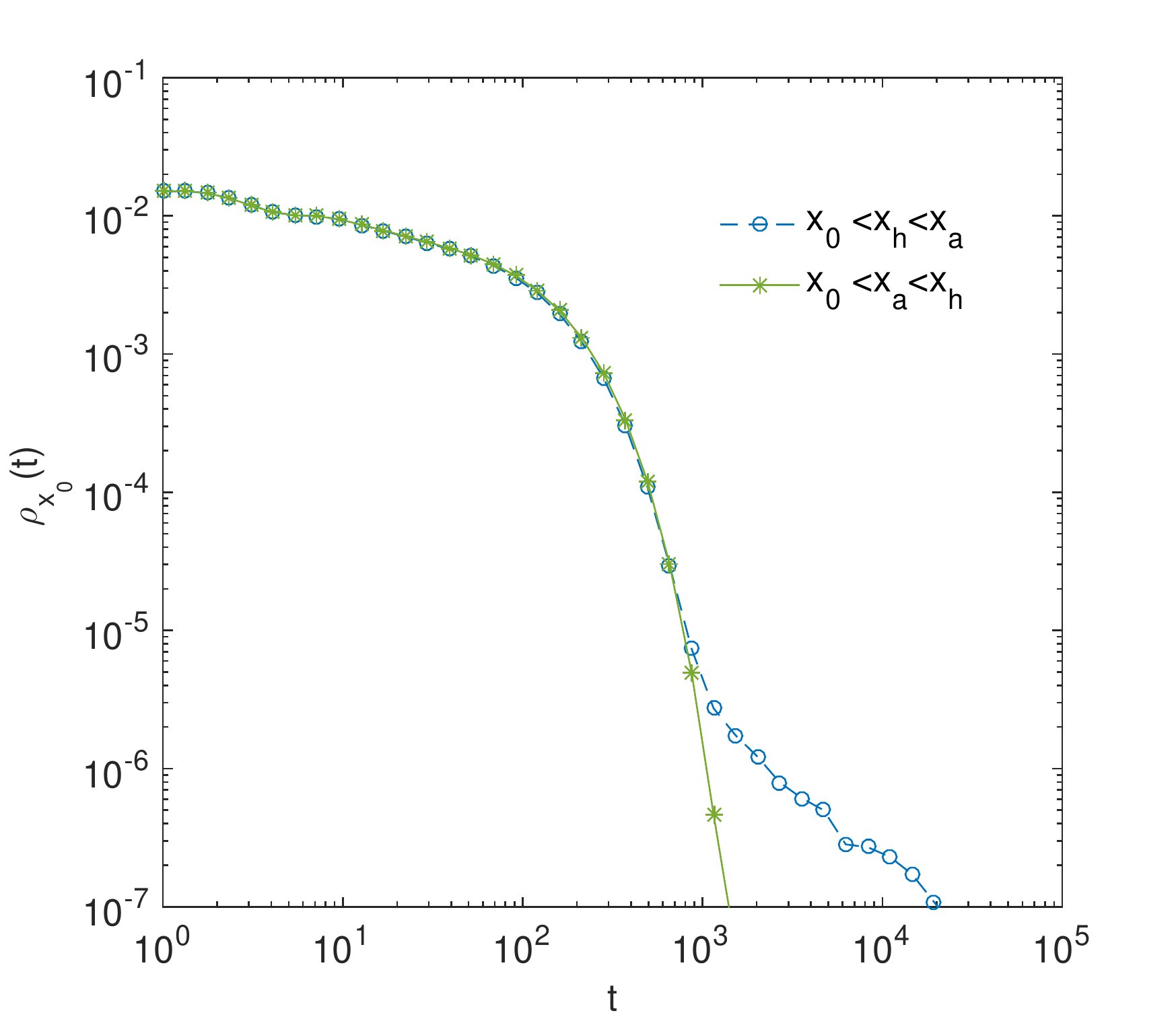}
	\caption{ FPT densities for HCTRW on SF network $G(N,m, m_0)$ for $N=100$, $m=5$, $m_0=2$ with two different positions of traps:
	 $x_h$ is placed  on a path $x_0<x_h<x_{a}$ (circles) and in another community of a network in respect to the starting node $x_0$, $x_0<x_{a}<x_h$  (stars). 
		Travel time distributions are  $\tilde{\psi}(s) = 1/(s\tau +1), \tau=1$, in a trap $\tilde{\psi}_{x_h}(s) = 1/(s^\alpha\tau^\alpha +1),$ $\alpha=0.1$.  
		$x_0$ and $x_{a}$ are fixed. 
	}	
	\label{fig_heter0}
\end{figure}

\begin{figure}
	\includegraphics[width=3.6in]{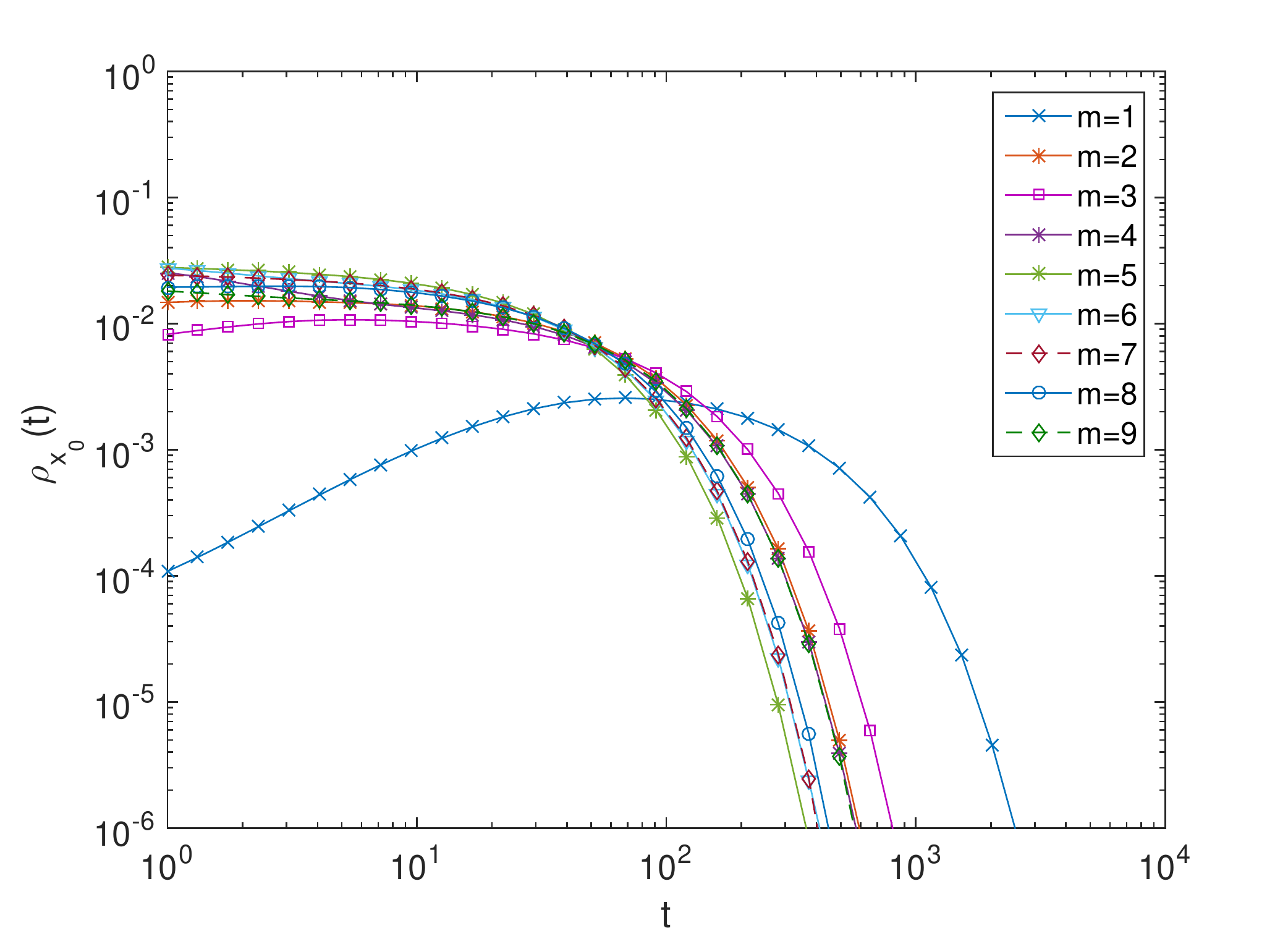}
	\caption{FPT density for SF networks with parameters $m_0=11$ and $m=1$ (tree SF network), $m\in[2,9]$ (non-tree SF network) with $N=100$ nodes. Travel time distribution for all the nodes is $\tilde{\psi}(s)=1/(s\tau+1),\tau=1$. $x_0$ and $x_{a}$ are randomly chosen in each SF network. 
	}	
	\label{fig_SFnets_diff_m}
\end{figure}

\subsection{Analytical insights onto the FPT density}
\label{subsec_FPT}

 Here we discuss the analytical calculations for the short-time and long-time regimes of the FPT density for a particular case ofs HCTRW model. 
Let us consider  a path of a random walk from $x_0$ to an absorbing node $x_{a}$ on a network with transition matrix $Q$. 
 If all travel time distributions are exponentials with the mean  $\tau$, then the probability of making the path of length $n$ is: 
 \begin{equation}
 P_{n}(t) = \frac{t^{n-1}}{\tau^n \Gamma(n)} e^{-t/\tau}.
 \label{eq_path}
 \end{equation}
 In the short-time regime the function $t^{n-1}$ dominates, while in the long-time regime the influence of the exponential $e^{-t/\tau}$ is much more profound.
 The total probability of arriving at $x_{a}$ from $x_0$ at time $t$ is: 
 \begin{equation}
 P_{x_0x_{a}}(t) = \sum_{n=1}^{\infty} (Q^n)_{x_0x_{abs}} P_n(t) = \\ \nonumber 
 \frac{e^{-t/\tau}}{\tau} \sum_{n=1}^{\infty} (Q^n)_{x_0x_{a}} \frac{(t/\tau)^{n-1}}{(n-1)!}.
\label{eq_propag_sp}
 \end{equation} 
The calculations of the propagator for a tree-like network can be simplified using the property that 
 the $n^{th}$ matrix power in the infinite series becomes zero matrix for $n\ge d$, where $d$ is the diameter of a network \cite{Petit2018}. 
 Variations of the shortest path $|x_0 - x_{a}|$ between the starting point $x_0$ with the fixed absorbing node $x_{a}$  affect 
 $(Q^n)_{x_0x_{a}}$. 

 The long-time behavior in finite  graphs, regardless of a network topology, has generic exponential tail \cite{Bollt}.
 However, this result is valid for the long-term behavior on structures without traps.  
 Let us consider a HCTRW model with induced structural or temporal perturbations, in which 
 all nodes have a finite-moment distribution $\psi(t)$  except one node $x_h$ with a heavy-tail distribution $\psi_{x_h}(t)$.
 Then the spectral properties of the generalized transition matrix $Q(t)$ are
 affected by structural ($Q\rightarrow Q_{pert}$)  and distributional  ($ \psi(t)$ for $x_h \rightarrow \psi_{x_h}(t)$) perturbations  \cite{GrebTupikina}.
 In the case of  distributional perturbations the graph is kept fixed with the same transition matrix $Q$ but the travel time distributions $\psi_{xx'}(t)$ are edge dependent and the generalized transition matrix $Q(t)$ changes. 
 In the case of structural perturbations, adding or removing links affects the matrix $Q$ itself, as the result, the stationary distribution changes. 
Both structural and distributional perturbations can be analyzed from the perspectives of temporal networks. In particular, 
 one can consider the behavior of a random walk in the HCTRW model as a simple random walk moving on continuously changing temporal network. In this temporal networks links are available in a certain time intervals, distributed according to some the probability distribution $e_{xx'}(t)$ for an edge $e_{xx'}$. Here we refer to the frameworks presented in \cite{Porter,CollizaTempo,Lambiotte2013}, where continuously evolving networks are studied.

 \subsection{Applications to real-world networks} \label{subsec_real}
 
\begin{figure}
 	\includegraphics[width=2.3in]{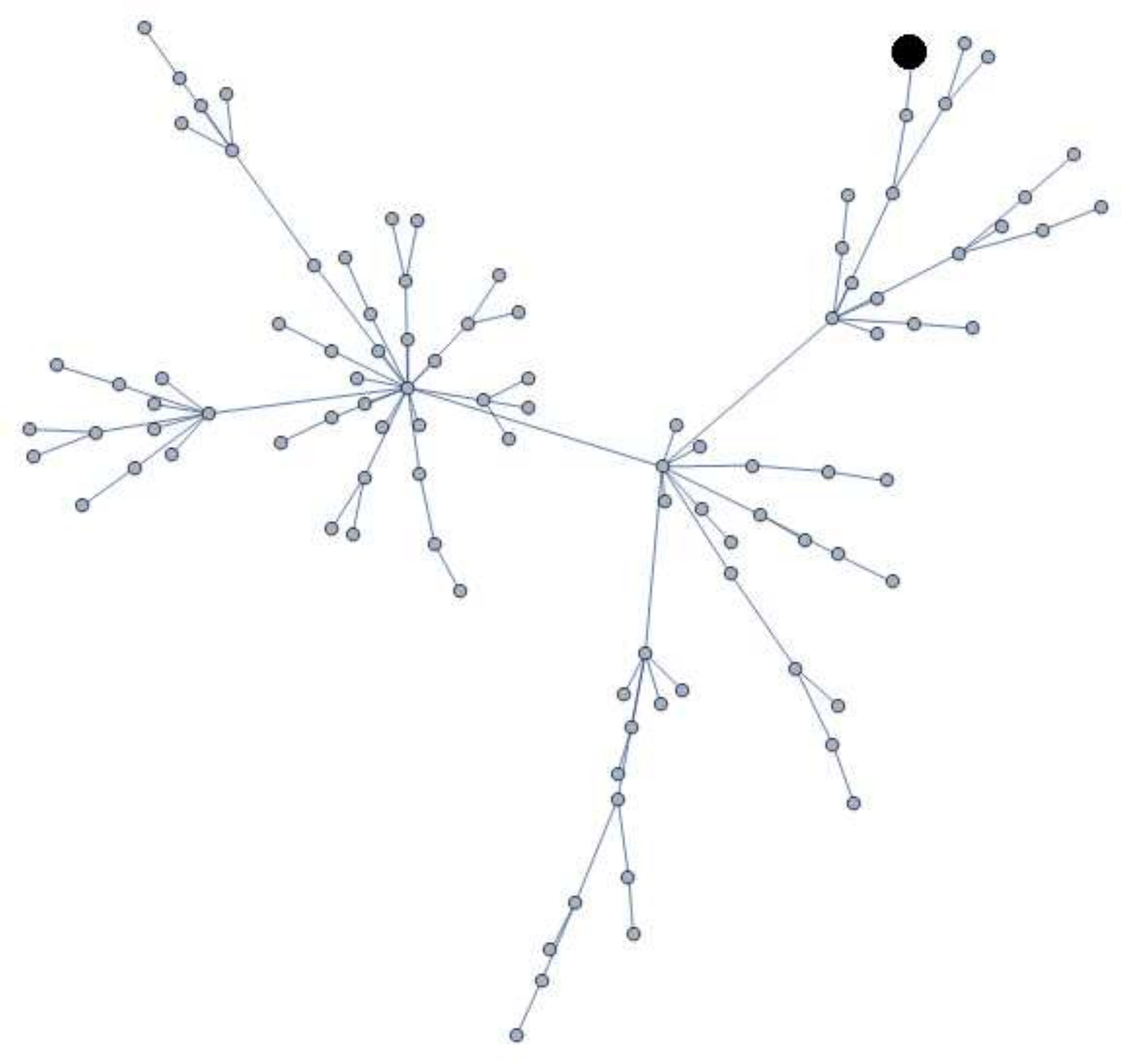}\\
\includegraphics[width=3.5in]{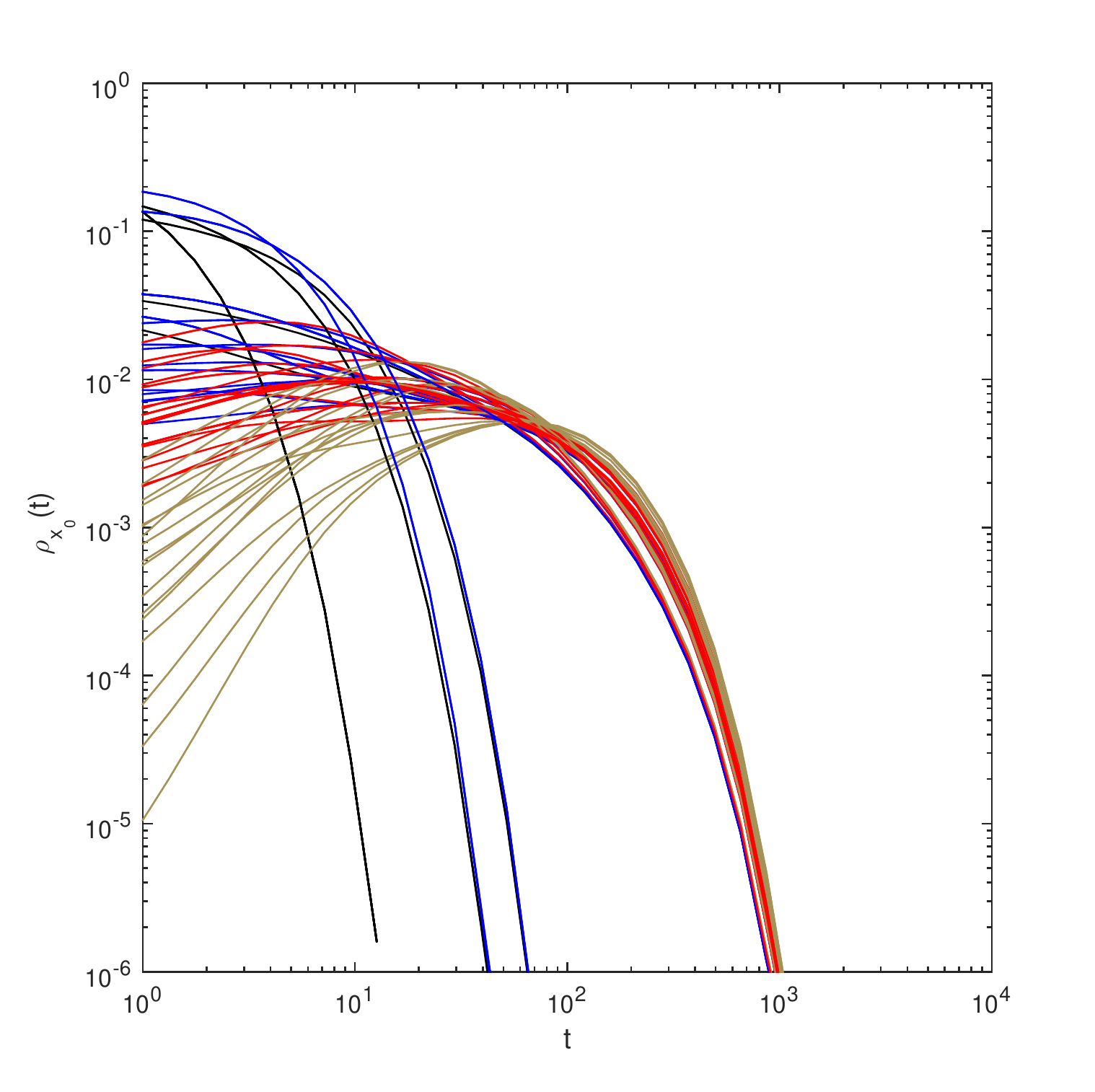}
\caption{(Top) Example of SF network with a fixed $x_{a}$ marked in black circle.
 		(Bottom)	FPT densities on SF network $G(N,m, m_0)$  $N=100$, $m=1$, $m_0=6$ 
 		for starting points $x_0$ and fixed $x_{a}$ such that: $|x_0-x_{a}|=1$ (black), $|x_0-x_{a}|=2$ (blue), $|x_0-x_{a}|=3$ (red) (FPT densities with $|x_0-x_{a}|>3$ are shown in grey). 
 		The travel time distribution is
 		$\tilde{\psi}(s) = 1/(s^\alpha\tau^\alpha +1)^2,  \alpha =1, \tau=1$, except in a fixed trap node $\alpha=0.5$. 
 	}	
 	\label{fig_hetermany}
 \end{figure}

 Let us give an illustrative example of the HCTRW  applications. 
 If a train station with a small number of connections  was shut down due to some technical problems, it would not significantly increase the number of additional trips that passengers would have to take to arrive at their final destinations. However, if a random perturbation affects a station with many connections or a connecting point between several clusters of stations, the average path length can increase dramatically.  
 In real transportation systems the  
 road occupancy is varying over time, as such, the link occupancy between two places $x,x'$ can be modeled by a random variable distributed with $\psi_{xx'}(t)$. Then the perturbation in the transportation liquidity (bursts) at some node can be modeled in the HCTRW as the trap $x_h$ with heavy-tail edge travel times. Links avoiding a trap correspond to additional transport connections, e.g. with buses etc.
 We consider a spreading entity (infection, rumour, flux of passengers, etc.), which can be modelled by a random walk behavior, although in the case of epidemics spreading the spreading quantity is not necessarily conserved. 
 We choose to analyze the London metro \cite{London}, 
 which shares some common features with other metro systems in the world, such as the average node degree of around $2.5$ inside the core of metro network \cite{RothBarthelemy}.  
 The London metro on Fig.~\ref{fig_London} (top) has $N=299$ stations (nowadays it has around $353$ stations and nearly $400$ edges). 
 The core of the London metro network, neglecting the radial stations outside the circle, exhibits  the small-world property.  
 We introduce a trap (perturbation) to London metro to see how this affects the transportation properties on the whole network.
 In Fig.~\ref{fig_London} we plot the FPT densities starting from different stations of the London metro and finishing at Piccadilly circus metro station (popular destination station). 
 Comparing  Fig.~\ref{fig_London} (bottom) with the FPT density for  SF and WS networks (Figs.~\ref{fig_hetermany}, \ref{fig_scalefreenontree}, \ref{fig_sw}) helps to get some insights. 
 The whole metro network of London is  separated into several groups of nodes according to the destination, in each group  the short-time regimes of $\rho_{x_0}(t)$ are similar. The long time regime is independent from the starting points (Fig.~\ref{fig_London}).

 \section{Discussions}
 \label{sec_disc}
 The properties of random walks  on various networks have been studied extensively  \cite{Sokolov,Krapivsky,Katsav}.
 Quite often the asymptotic analysis and the averaged characteristics of diffusive transport and random walk dynamics, such as mean return time and mean first-passage time, are  calculated \cite{Bollt,Brockman}. 
 Here we focused on studying the probability distribution of the first-passage time for random walk on complex networks  with heterogeneities.

 \begin{figure}
 	\includegraphics[width=2.3in]{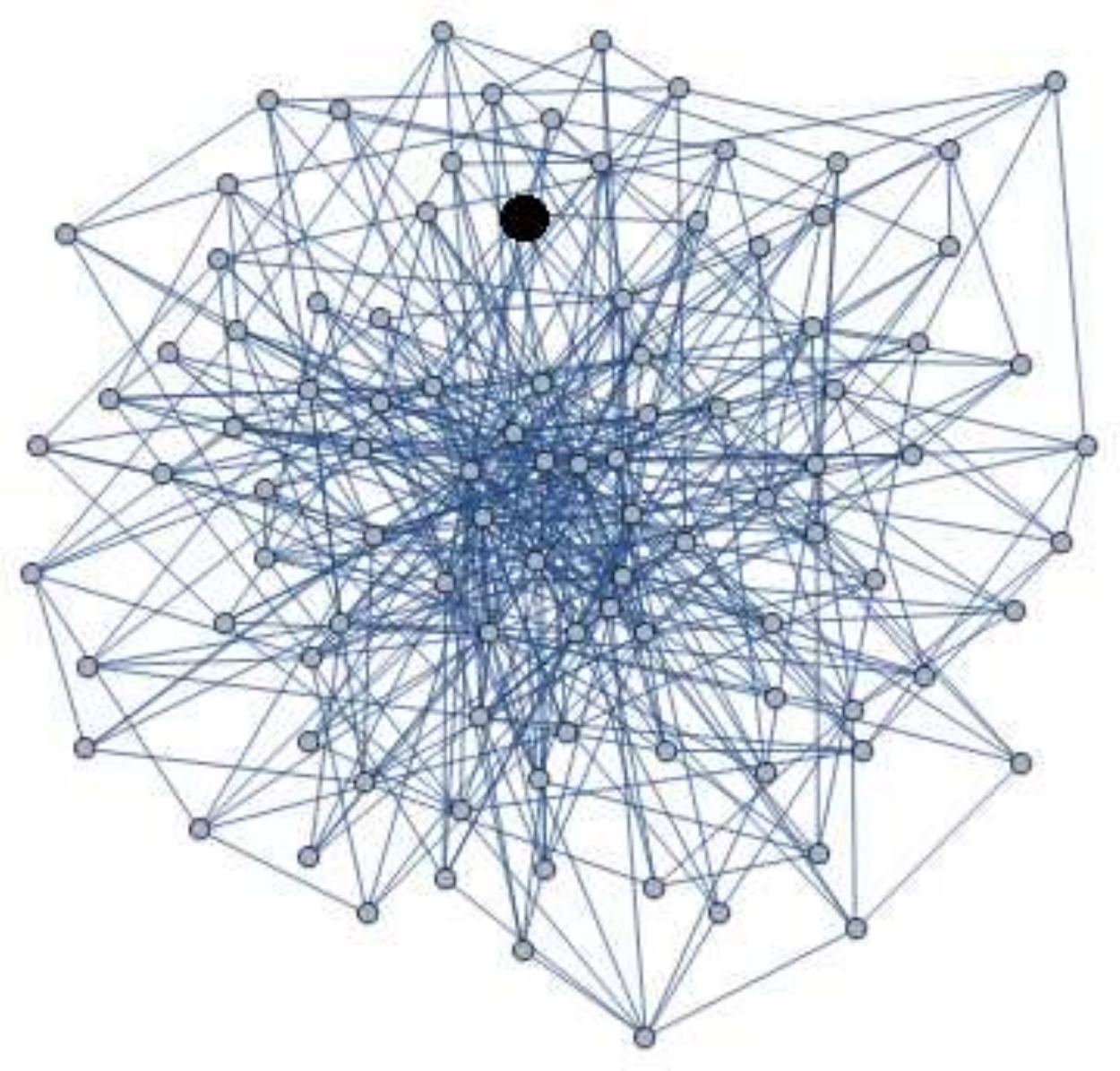}
 	\\
 	\includegraphics[width=3.6in]{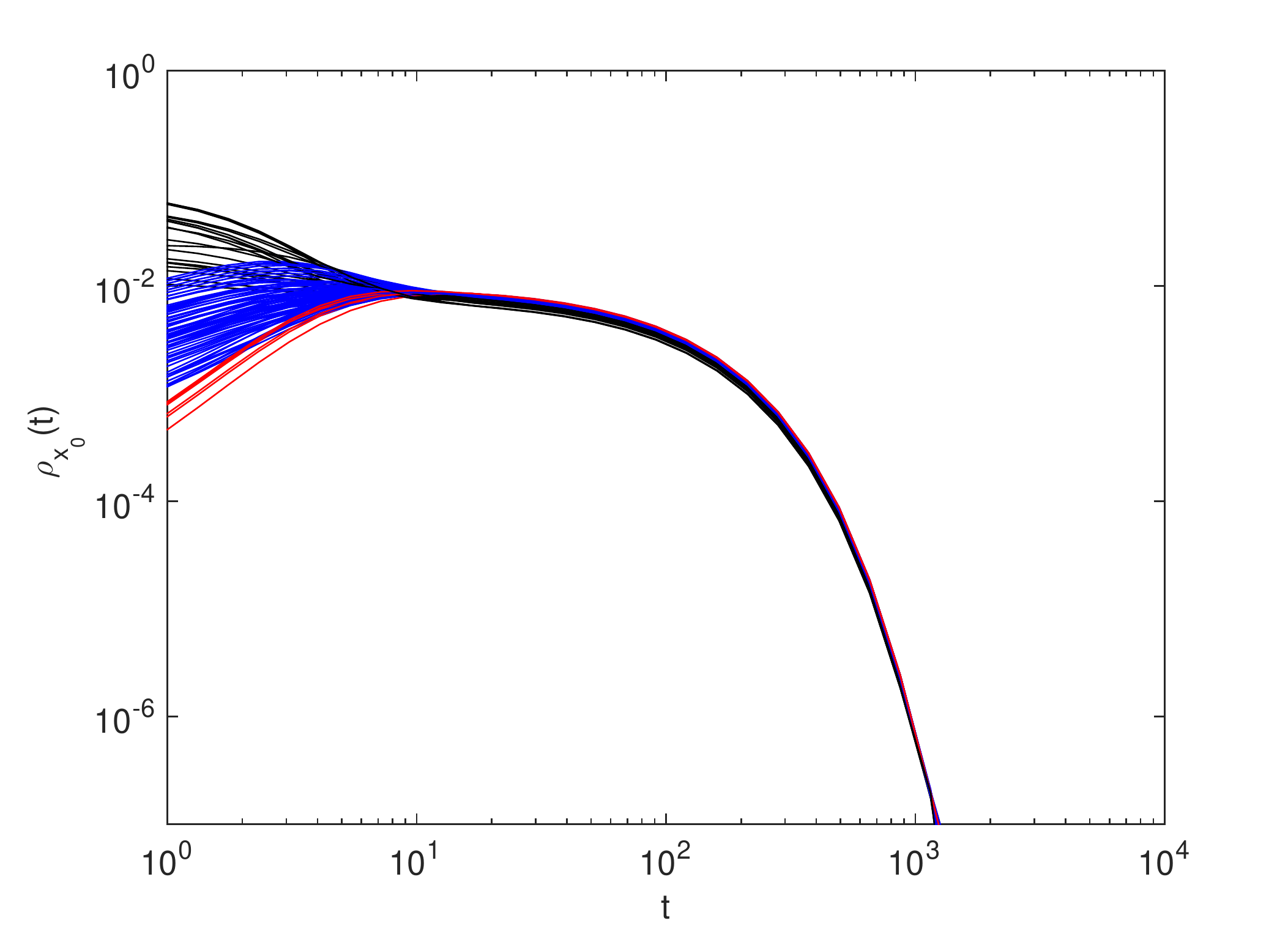}
 	\caption{
 		(Top) Example of SF network with fixed $x_{a}$ marked by black circle.
 		(Bottom)	FPT densities on SF network $G(N,m, m_0)$  $N=100$, $m=5$, $m_0=6$ 
 		for different starting points $x_0$ and fixed $x_{a}$: $|x_0-x_{a}|=1$ (black),  $|x_0-x_{a}|=2$ (blue),  $|x_0-x_{a}|=3$ (red). 
 		The travel time distribution is
 		$\tilde{\psi}(s) = 1/(s^\alpha\tau^\alpha +1)^2, \alpha=1, \tau=1$, except in a fixed trap node $\alpha=0.5$. 
 	}	
 	\label{fig_scalefreenontree}
 \end{figure}

 We identified three regimes of the FPT density (Fig.~\ref{fig_fpt_scheme}): short-time, intermediate-time and long-time regimes.
This 
 scheme also  
 demonstrates the observation about how the FPT density is affected 
 by different types of perturbations.
 In particular, local structural perturbations (such as an inclusion of links avoiding $x_h$) 
 mainly affect the short-time regime of the FPT density.
 The global structural perturbations (local perturbations of the network structure are accumulated and affect the network topology globally) 
 generally have an effect on	the intermediate regime.
 Finally,  the distributional perturbations (when the travel times $\psi_{xx'}(t)$ are link-dependent) are mainly affecting the long-time regimes of the FPT density.
 One can also consider the interplay of different types of perturbations but this goes beyond the scope of this paper. 
 We start with the least studied short-time regime \cite{Bollt}.

 The short-time regime is strongly affected by the geometric network properties, such as the distance between $x_0$ and $x_{a}$, which can be calculated, 
 as the minimal power of the adjacency
 matrix of a network with nonzero element $A_{x_0x_{a}}$:  $A^{L-1}_{x_0x_{a}} =0, {A^L}_{x_0x_{a}}\neq 0.$ 
 As we saw from the numerical computations of $\rho_{x_0}(t)$ (subsection \ref{sec_trees}), a local perturbation induced by an avoiding link strongly affects the short-time regime (Fig.~\ref{fig_heter_avoid_fract} (bottom)). In other words, local changes of a network structure affect the distribution of the shortest paths, which in turn affects the left tail of the FPT density (Figs.~\ref{fig_hetermany},\ref{fig_scalefreenontree}).
 In the same vain, the FPT properties 
 are strongly affected by the metrics of a graph and not just by the node degrees, in the contrary to the common belief that general transport properties of networks depend exclusively on node degrees.  
At the same time, the non-linear network measure distribution is related to dynamical properties in a rather complex way  and should be analyzed separately.
 
 \begin{figure}
 	\includegraphics[width=2.5in]{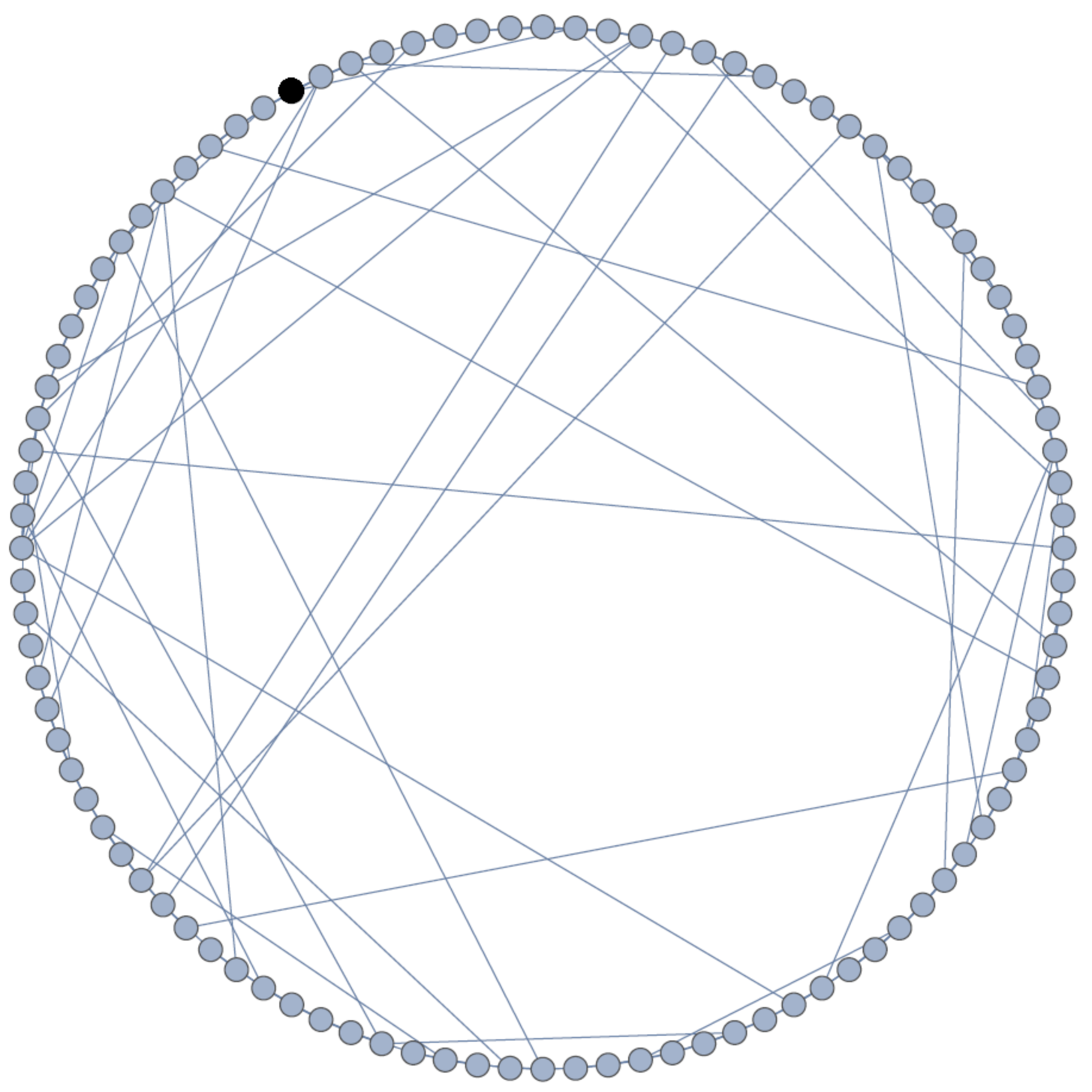}
 	\\
 	\includegraphics[width=3.2in]{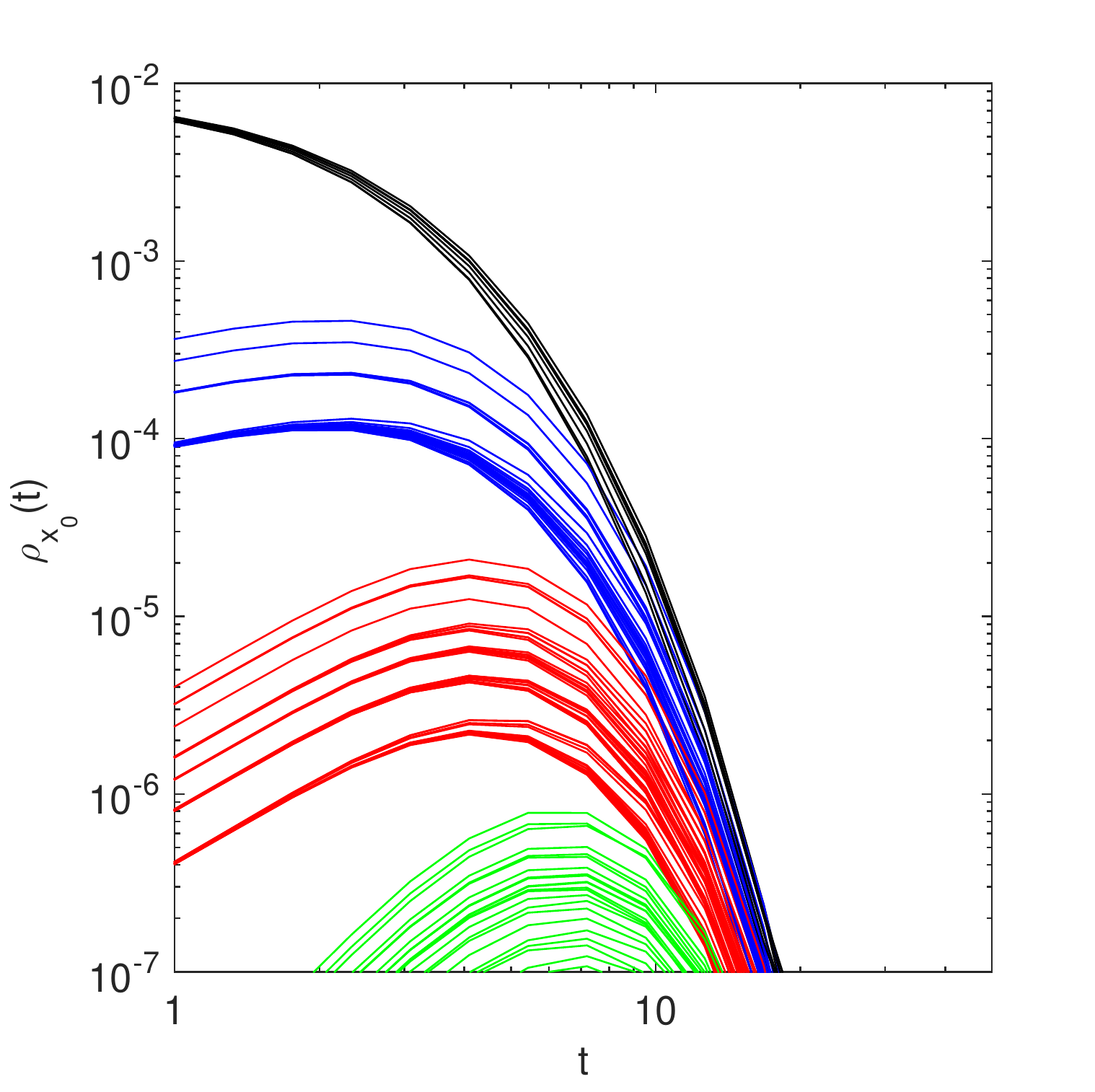}
 	\caption{ (Top) Example of WS network model with $\beta =0.2, k=8, N=100$.  (Bottom) FPT density for this network
 		with homogeneous travel time distributions $\tilde{\psi}(s) = 1/(s\tau +1)$ for different initial points $x_0$, $x_{a}=1$. Different colors correspond to different shortest paths length between different $x_0$ and fixed $x_{a}$: $|x_0-x_{a}|=1$ (black), $|x_0-x_{a}|=2$ (blue), $|x_0-x_{a}|=3$ (red),$|x_0-x_{a}|=4$ (green).   }	
 	\label{fig_sw}
 \end{figure}
 
 The intermediate-time regime of the FPT density
 mainly depends on the global topological features of a network (i.e. small-world property, loopless structure). 
 For some networks the intermediate regime is less pronounced than for others (compare Figs.~\ref{fig_hetermany} and \ref{fig_scalefreenontree}). 
 For instance, for WS model (Fig.~\ref{fig_sw}), the intermediate regime between the most probable first-passage time, $t_{mp}$, and beginning of a decrease of the FPT density is not so well pronounced.  
 For the networks without traps 
 changing 
 the shortest path between $x_0$ and $x_{a}$ shifts the maximum of the FPT. 
 This resembles the properties of diffusion in  continuous domains \cite{Godec}, where $t_{mp}$ is proportional to a  distance between the starting point and the target.   
 Another important characteristics of the intermediate regime is 
 the presence of a plateau, like the intermediate regime at  $t\in[10^1, 10^3]$ for SF networks with  $m=5, m_0=6$ (Fig.~\ref{fig_scalefreenontree}), which is a general feature present also for FPT in continuous  domains \cite{Greb2018}.
 
 \begin{figure}	
 	\includegraphics[width=2.5in]{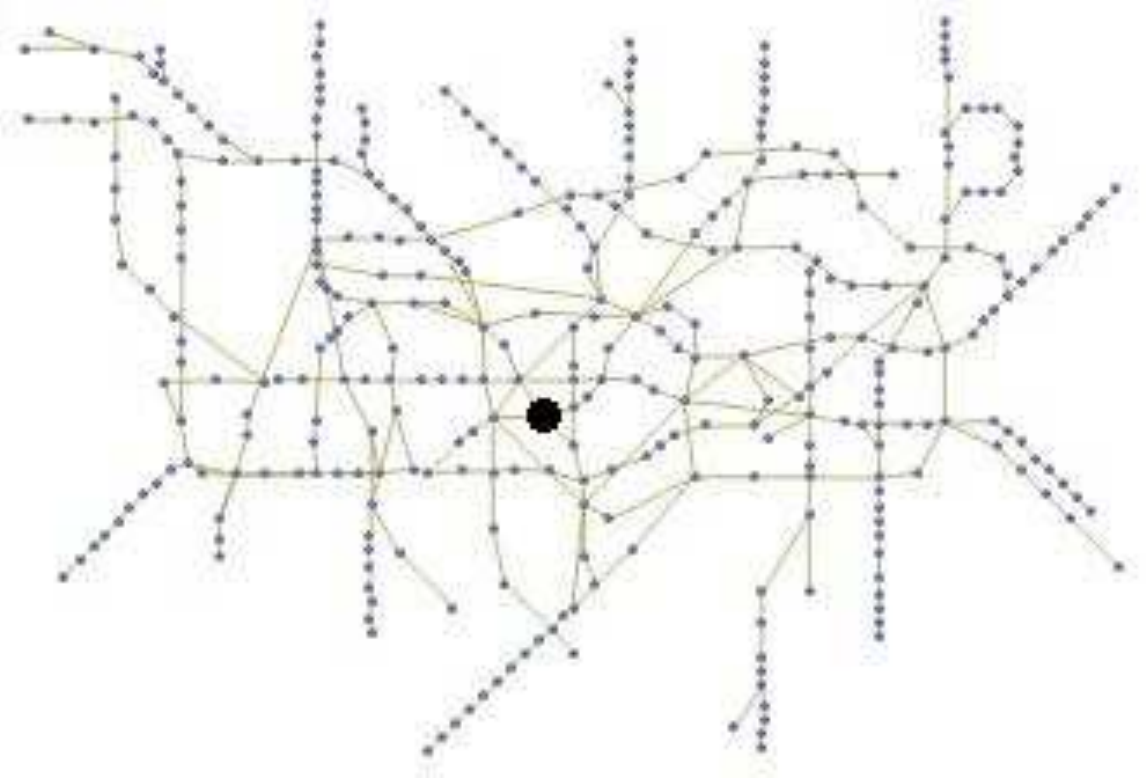}
 	\\
 	\includegraphics[width=3.3 in]{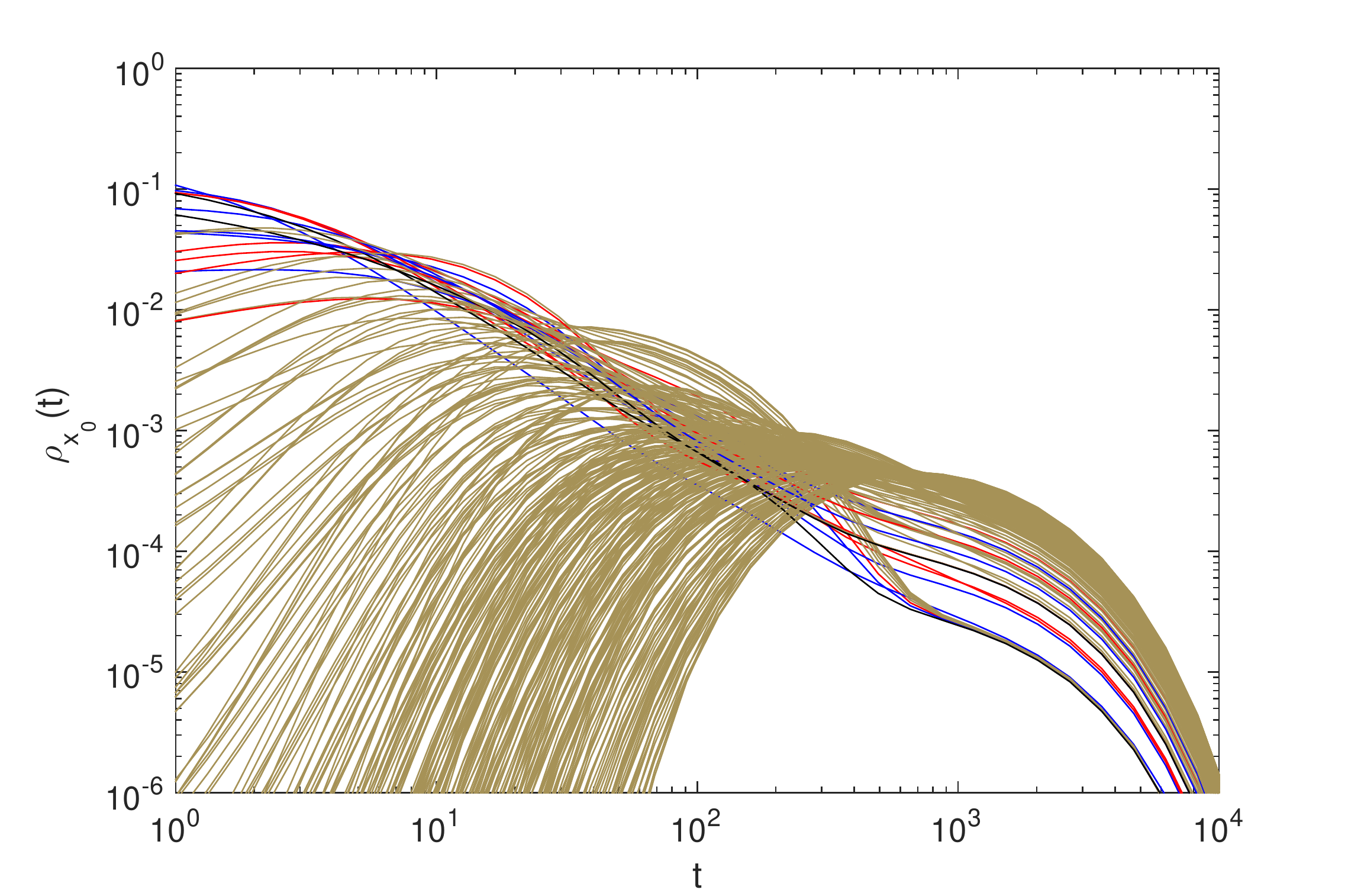}%
 	\caption{(Top) The  London metro map with destination Piccadilly station $x_{a}$ highlighted in black. (Bottom)
 		FPT density for the London metro with  travel time distributions $\tilde{\psi}(s) = 1/(s\tau +1)$ except at a trap node $x_h$ at which $\tilde{\psi}_{x_h}(s) = 1/(s^\alpha\tau^\alpha +1), \alpha=0.5$. For each starting stations $x_0$ the FPT density is colored  according to the distance from $x_0$ to $x_{a}$ (Piccadilly station): $|x_0-x_{a}|=1$ (black), $|x_0-x_{a}|=2$ (blue), $|x_0-x_{a}|=3$ (red) (FPT densities with $|x_0-x_{a}|>3$ are shown in grey). }	
 	\label{fig_London}
 \end{figure}

 Finally, the long-time regime is largely influenced by temporal heterogeneities, i.e. the trap nodes $x_h$ with a heavy tail distribution of travel time $\psi_{x_hx'}$. 
 In particular, it was identified 
 that for simple random walks MFPT (mean first-passage time) depicts some long-time properties of the dynamics. However, MFPT neglects some important information about the process, 
 significantly overestimating the scales of the most-probable first-passage time \cite{Godec,Greb2018}.
 At the same time it is known that for  
 dynamics of a particle diffusing in a continuous domain the MFPT  is of order  $L^2/D$, where $L$ is the domain size,  $D$ is the diffusion coefficient \cite{Singer}. 
 In the continuous case the most probable FPT is shown to strongly depend on the starting position and is almost independent of the target properties  \cite{Greb2018}.   
 How the dynamical properties are changing when distributional perturbation at $x_h$ is introduced?
 The influence of $x_h$ on the long-time regime is illustrated on Fig.~\ref{fig_heter0}, where placing $x_h$ in different communities of a network changes the right tail of the FPT density.  
 Note that the trap does not affect the geometric length of the path between $x_0$ and $x_{a}$, but may affect the shortest-in-time path between nodes. 
 Hence even if there are several paths between nodes
 there is a non-zero probability for a random walk to reach $x_h$ at long times, which then may affect the right tail of the distribution (Fig.~\ref{fig_heter_avoid}). 
 To complete our discussions on this point, we recall that the properties in the long-time regime of random walks on networks without heterogeneities have been proven to depend on the dimension of a random walk  \cite{Procacia}.
 Moreover, in the case of the compact random walk exploration, e.g. HCTRW on Viscek fractals with spectral dimension $d_s<2$ (Fig.~\ref{fig_heter_avoid_fract}), the asymptotic properties of the FPT density  do not depend on a degree of a target node and have the power law decay $t^{-2d_s-1-\alpha}$.

 \begin{figure}	
 	\includegraphics[width=3.5in]{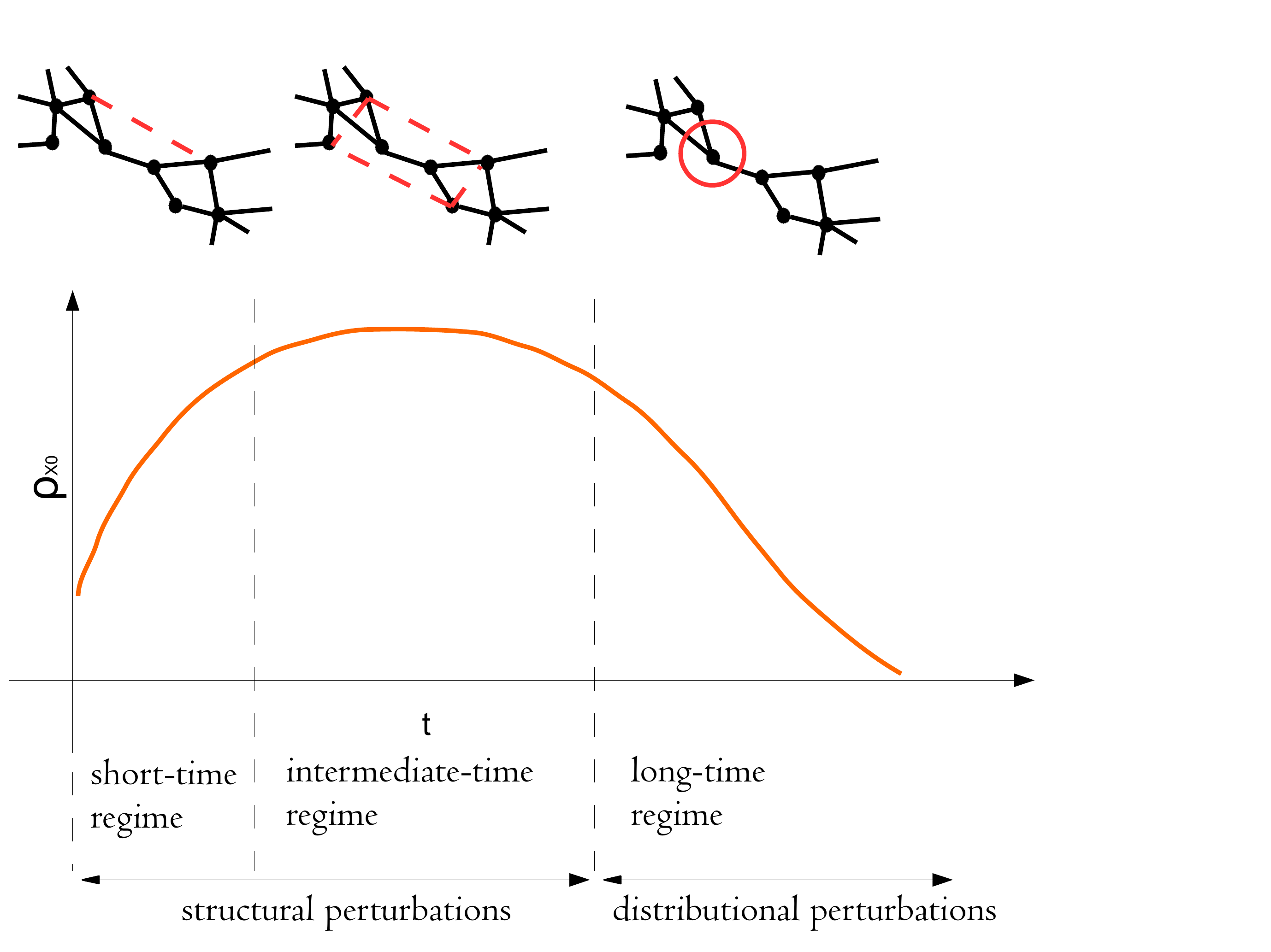}
 	\caption{ The schematic representation of the short-time, intermediate-time and long-time regimes of the FPT density on various networks.  }	
 	\label{fig_fpt_scheme}
 \end{figure}

 \section{Conclusions}
 \label{sec_conc}
 
 In this article we considered the HCTRW framework for studying the feature-rich networks, in particular, we studied effects of heterogeneity on random walk dynamics on several types of networks.  
 We explored FPT densities on regular and irregular structures, such as regular fractals, random Scale-Free and Watts-Strogatz network models. 
 The comparison of regular versus irregular structures 
 allowed us 
 to refine some properties of the FPT densities on complex networks. 
 In particular we analyzed effects of structural and distributional heterogeneities using the first-passage time 
 as one of the key indicators of how fast information diffuses in a given system
 \cite{Brockman,Redner2002}. 
 Heterogeneities are   encoded in the generalized transition matrix $Q(t)$ \cite{GrebTupikina}, which affects its  spectral properties and as the result the dynamical properties of processes on such networks.
 The influence of the structural and distributional perturbations of network structure on FPT density is summarized in Fig.~\ref{fig_fpt_scheme}. 
 The local and global structural perturbations mainly influence the short-time and intermediate-time regimes. The distributional  perturbations generally affect the long-time regime. 
 Notably, not only the topology of the network but also temporal discrepancy encoded in the distributional heterogeneities  $\psi_{xx'}(t)$
 can significantly alter the behavior of random walk on a network. 
 
 In particular, 
 we found that the short-time regime depends on the 
 distribution of distances between 
 starting and absorbing  nodes in a network. 
 The existence of loops generally influences the dynamical properties of random walks on these graphs \cite{Dorogovtsev}. 
 For the HCTRW model the changes of global network structure influence the intermediate-time regime of the FPT density, e.g. SF network with or without cycles has different intermediate regime duration (section \ref{sec_num_res}).  
 In general, analysis of the short-time and intermediate-time regimes allow us
 to characterize transport properties of networks even when the first moment of the distribution is infinite  \cite{Hernandez1990}. 
 
 To summarize, the FPT density provides us with alternative characteristics of network  topology in addition to known static measures, such as degree or closeness network measures \cite{Barthelemy}. 
 The average number of distinct sites, visited by HCTRW and other observables derived from the FPT densities can be a quantity with the practical relevance for charaterization of spreading processes.

   The presented analysis of random walk dynamics can be further explored on real-world networks, 
   as it has been done with the CTRW approach for analysis of human travel laws in the continuous domains \cite{BrockmanPopul}. 
 We applied the HCTRW model on a network graph of the London metro. 
 The next step would be to compare the results from HCTRW simulations, Fig.~\ref{fig_London}, with the original data of travelling passengers on real-world networks \cite{datastanf}.
  The transportation network is 
  the network type, to which the HCTRW framework can be further applied, taking into account their nontrivial stochastic nature. 
   Moreover, the HCTRW framework can be used in order to navigate in the network using detection of the relevant search strategies, e.g. placing heterogeneities in different parts of a network. Using the first-passage time observables one can further design network measures 
    in order to characterize
    transport efficiency and spreading. 
The straightforward way to characterize the transport and search efficiency is to use the concept of survivability as a measure of reachability of some quantity in a network, which we plan to explore in the future. 
We expect the framework presented in this paper to  broaden the scopes of exploration of the feature-rich networks.


 \section{Declaration}
\subsection{Availability of data and material}
The datasets used and/or analysed during the current study are available from the corresponding author on reasonable request.

\subsection{Competing interests}
The authors declare that they have no competing interests.

\subsection{Funding}
The authors acknowledge the support under Grant No. ANR-13-JSV5-0006-01 of the French National Research Agency. L.T. acknowledges support of Centre of Research and Interdisciplinarity, Paris Descartes University, France.

\subsection{Authors' contributions}
LT performed numerical simulations, analyzed and interpreted the analytical and numerical results of the HCTRW model on various types of networks. DG analyzed, interpreted the analytical and numerical results. Both authors contributed in writing the manuscript. Both authors read and approved the final manuscript.

\subsection{Acknowledgements.} 
 L.T. thanks M. Dolgushev for exchange of materials about fractal networks and P. Holme for discussions about network measures for spreading.  
 

\section{ List of abbreviations}
\emph{HCTRW} model - Heterogeneous Continuous Time Random Walk model,
\\
\emph{FPT} density - First-Passage Time density, 
\\
\emph{SF} network - Scale-Free network, 
\\
\emph{WS} network - Watts-Strogatz network.

\section{FIGURE LEGEND}
Figures:
\\
\textbf{Fig.1} FPT density on a chain graph with $N=100$ nodes,   $\tilde{\psi}(s) = 1/(s\tau +1), \tau=1$ in all nodes except the trap nodes either at $x_h = 25$, or at $x_h = 75$, at which $\tilde{\psi}_{x_h}(s) = 1/(s^\alpha\tau^\alpha +1), \alpha=0.5$,  $x_0=50$ and $x_{a}=1$. The additional link  is placed around the trap $x_h=25$. 
\\
\textbf{Fig.2}
(Top)
Vicsek fractal $G_{V33}$ with $N=64$ nodes.  The starting node $x_0$ is in the center (red circle), the absorbing node $x_{a}$ is one of the dead-ends (black circle), traps $x_h$ (empty circles). (Bottom)
FPT densities on this Vicsek fractal with   $\tilde{\psi}(s) = 1/(s\tau +1), \tau=1$ in all nodes except $x_h$, at which  $\tilde{\psi}(s) = 1/(s^\alpha\tau^\alpha +1), \alpha=0.5$. 
The trap $x_h$ is placed either on the shortest path between $x_0$ and $x_{a}$ ($x_h=14$), or outside ($x_h=54$). We also consider the case with an
additional link avoiding the trap $x_h=14$, which corresponds to the transition matrix with a local perturbation.
\\
\textbf{Fig.3}
FPT densities for HCTRW on SF network $G(N,m, m_0)$ for $N=100$, $m=5$, $m_0=2$ with two different positions of traps:
$x_h$ is placed  on a path $x_0<x_h<x_{a}$ (circles) and in another community of a network in respect to the starting node $x_0$, $x_0<x_{a}<x_h$  (stars). 
Travel time distributions are  $\tilde{\psi}(s) = 1/(s\tau +1), \tau=1$, in a trap $\tilde{\psi}_{x_h}(s) = 1/(s^\alpha\tau^\alpha +1),$ $\alpha=0.1$.  
$x_0$ and $x_{a}$ are fixed.
\\
\textbf{Fig.4}
FPT density for SF networks with parameters $m_0=11$ and $m=1$ (tree SF network), $m\in[2,9]$ (non-tree SF network) with $N=100$ nodes. Travel time distribution for all the nodes is $\tilde{\psi}(s)=1/(s\tau+1),\tau=1$. $x_0$ and $x_{a}$ are randomly chosen in each SF network. 
\\
\textbf{Fig.5}
(Top) Example of SF network with a fixed $x_{a}$ marked in black circle.
(Bottom)	FPT densities on SF network $G(N,m, m_0)$  $N=100$, $m=1$, $m_0=6$ 
for starting points $x_0$ and fixed $x_{a}$ such that: $|x_0-x_{a}|=1$ (black), $|x_0-x_{a}|=2$ (blue), $|x_0-x_{a}|=3$ (red) (FPT densities with $|x_0-x_{a}|>3$ are shown in grey). 
The travel time distribution is
$\tilde{\psi}(s) = 1/(s^\alpha\tau^\alpha +1)^2,  \alpha =1, \tau=1$, except in a fixed trap node $\alpha=0.5$. 
\\
\textbf{Fig.6}
(Top) Example of SF network with fixed $x_{a}$ marked by black circle.
(Bottom)	FPT densities on SF network $G(N,m, m_0)$  $N=100$, $m=5$, $m_0=6$ 
for different starting points $x_0$ and fixed $x_{a}$: $|x_0-x_{a}|=1$ (black),  $|x_0-x_{a}|=2$ (blue),  $|x_0-x_{a}|=3$ (red). 
The travel time distribution is
$\tilde{\psi}(s) = 1/(s^\alpha\tau^\alpha +1)^2, \alpha=1, \tau=1$, except in a fixed trap node $\alpha=0.5$. 
\\
\textbf{Fig.7}
(Top) Example of WS network model with $\beta =0.2, k=8, N=100$.  (Bottom) FPT density for this network
with homogeneous travel time distributions $\tilde{\psi}(s) = 1/(s\tau +1)$ for different initial points $x_0$, $x_{a}=1$. Different colors correspond to different shortest paths length between different $x_0$ and fixed $x_{a}$: $|x_0-x_{a}|=1$ (black), $|x_0-x_{a}|=2$ (blue), $|x_0-x_{a}|=3$ (red),$|x_0-x_{a}|=4$ (green). 
\\
\textbf{Fig.8}
(Top) The  London metro map with destination Piccadilly station $x_{a}$ highlighted in black. (Bottom)
FPT density for the London metro with  travel time distributions $\tilde{\psi}(s) = 1/(s\tau +1)$ except at a trap node $x_h$ at which $\tilde{\psi}_{x_h}(s) = 1/(s^\alpha\tau^\alpha +1), \alpha=0.5$. For each starting stations $x_0$ the FPT density is colored  according to the distance from $x_0$ to $x_{a}$ (Piccadilly station): $|x_0-x_{a}|=1$ (black), $|x_0-x_{a}|=2$ (blue), $|x_0-x_{a}|=3$ (red) (FPT densities with $|x_0-x_{a}|>3$ are shown in grey).
\\
\textbf{Fig.9}
The schematic representation of the short-time, intermediate-time and long-time regimes of the FPT density on various networks.  
\end{document}